# Quantum Proteomics

Fabio Pichierri[a,b*]

[a]*G-COE Laboratory, Department of Applied Chemistry, Graduate School of Engineering,*

*Tohoku University, Aoba-yama 6-6-07, Sendai 980-8579, Japan*

[b]*Quantum Proteomics Initiative (QPI)*

[Dated: July 29, 2011]

## Abstract

We put forward the idea of establishing a novel interdisciplinary field of research at the interface between quantum mechanics and proteomics. The new field, called *quantum proteomics,* is defined as *the large-scale study of the electronic structure of the proteins that define an organism's proteome*. The electronic structure of proteins is unveiled with the aid of linear-scaling quantum mechanical calculations. Such calculations provide information about the energy levels of the proteins, the charges of their amino acid side chains, their electrostatic potentials and permanent dipole moments ($\mu$). Since the magnitude of the electric dipole moment of any protein is not null ($\mu \neq 0$ Debye), the dipole moment can be employed to characterize the electronic structure of each protein that belongs to an organism's proteome. As an example, we investigate six proteins from the thermophilic bacterium *Methanobacterium thermoautotrophicum* (Mth) whose atomic structures were characterized by solution NMR spectroscopy.

[*] Corresponding author. Tel. & Fax: +81-22-795-4132

*E-mail address:* fabio@che.tohoku.ac.jp (F. Pichierri)

**Introduction**

Proteomics, the field of research concerned with the large-scale study of proteins, gained rapid momentum soon after the completion of the human genome project (HGP) in the early 2001 [1-3]. The identification of about 30,000 genes by HGP paved the way toward the identification and study of the proteins that are expressed by those genes, i.e. the *proteome* which is defined as the complement of the genome. According to the Human Proteome Project (HPP), the number of protein-coding genes is estimated being close to 20,300 [4]. As of July 2011, the atomic coordinates of more than 70,000 protein structures (from different organisms) have been determined by X-ray crystallography, NMR spectroscopy, and electron microscopy and their coordinates deposited into the Protein Data Bank (PDB) [5]. The determination of the structure of all proteins defining the human proteome as well as any other organism's proteome is therefore still a very far sight but the large number of proteomics projects that have been started in many countries coupled to technological advances in crystallization and structure determination is speeding up the analysis of the human proteome. Besides the characterization of protein structures, important issues in proteomics research are represented by the study of protein-protein interactions, which define the so called *interactome*, and protein phosphorylation (*phospho-proteomics*) which is concerned to cell signaling [6]. Mass spectrometry (MS) is an important tool for proteomics research that, when coupled to computational studies (often grouped under the name of *computational proteomics*), is likely to advance the field even without knowledge of the 3D structures of proteins [7].

On a different front, theoretical physicists are proposing that complex biological phenomena, mostly involving the interaction among different proteins and other biomolecules, are based on quantum mechanics (QM) [8]. For example, photosynthesis in plants and bacteria and the information processing in microtubules, which is responsible for brain function and emergent properties like consciousness, are suggested to occur as a result of QM being operative within biological environments [9-13]. Several theories and ideas in this field known as *quantum biology* [14] are still hotly debated and we do not enter into their merit here. Theoretical and computational chemists, on the other hand, started to study the electronic structure of whole protein molecules with the aid of quantum mechanical calculations (*quantum biochemistry*) [15]. The author, for instance, has recently investigated two membrane proteins, the KcsA potassium channel [16] and the LH2 complex of purple bacteria [17], both of which are characterized by giant, permanent dipole moments which indicate a strong polarization of electronic charge with respect to the membrane's mean plane.

All the above research indicates that there exist strong interdisciplinary research efforts that are ongoing at the interface between biology and the neighboring fields of physics and chemistry. Following these efforts toward understanding biological systems at a quantitative level, here we put forward the idea of establishing a novel interdisciplinary field of research termed *quantum proteomics*. We define quantum proteomics (QP) as *the large-scale study of the electronic structure of the proteins that define an* *organism's proteome.* The electronic structure of six proteins from *Methanobacterium thermoautotrophicum* (Mth) [18] is investigated with the aid of linear-scaling quantum mechanical calculations. Such calculations provide information about the energy levels of the proteins, the charges of their amino acid side chains, their electrostatic potentials and permanent dipole moments ($\mu$). Since for any protein $\mu \neq 0$ Debye, the electric dipole moment is here employed to characterize each protein of the Mth proteome.

**Methods**

Electronic structure calculations were performed with the MOPAC2009 software package [19]. This software implements MOZYME, a linear-scaling routine which employs the Lewis structure of the protein for the initial guess of the density matrix and localized orbitals for the self-consistent field (SCF) calculations [20], as shown schematically in Figure 1. The QM calculations employed the new semiempirical PM6 parameters of Stewart [21] which have been recently applied to the investigation of different proteins [22]. The effect of the solvent (water) environment was included with the Conductor-like Screening Model (COSMO) [23] by setting the dielectric constant of the medium to the value $\varepsilon=78.4$.

The geometries of the six proteins investigated herein were obtained from the Protein Databank (PDB) [5]: Mth0637 (PDB id 1JRM), Mth0865 (PDB id 1IIO), Mth0895 (PDB id 1ILO), Mth1692 (PDB id 1JCU), Mth1743 (PDB id 1RYJ), and Mth1880 (PDB id 1IQS). Because the structures of these proteins were determined by solution NMR spectroscopy [18], no further geometry optimization was carried out. Also, since all the above PDB structures contain hydrogen atoms, the protonation state of the experimental NMR structure determination was adopted in this study. For Mth1880 the average NMR structure was employed while for the other proteins the first conformer (denoted to as model 1 inside the corresponding PDB file) was employed. One should notice that the dipole moment of a charged protein can be computed with respect to its center of mass and by employing an accelerating frame of reference, as discussed by Stewart [19].

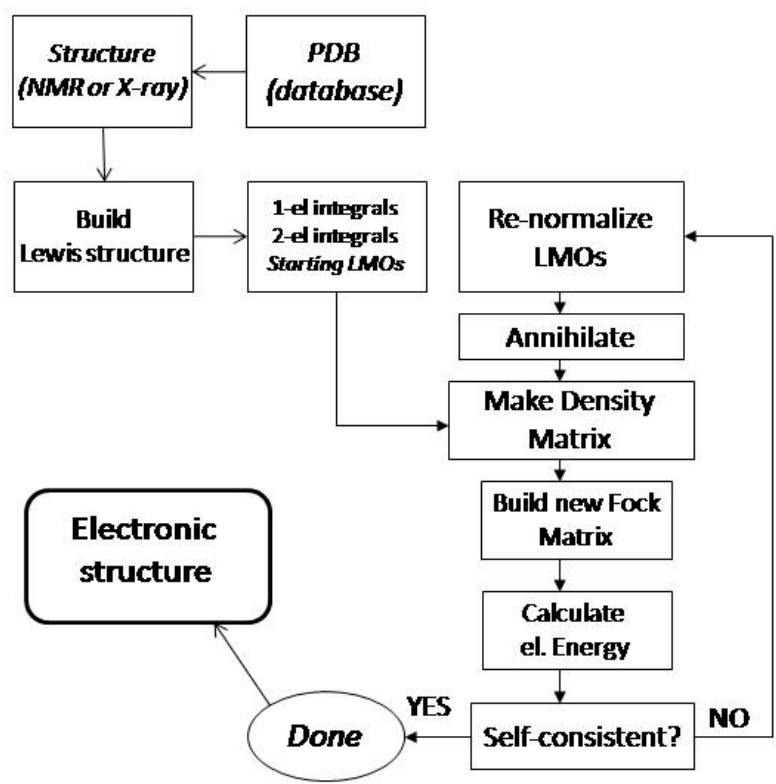

**Figure 1.** Flow chart showing the steps necessary to obtain the electronic structure of the protein starting from its 3D atomic coordinates downloaded from PDB. Modified and adapted from [20].

**Results and discussion**

Table 1 reports the results of our quantum mechanical calculations performed on six Mth proteins whose 3D structures were characterized by NMR spectroscopy [18]. These proteins differ in size as indicated by the number of atoms (NA parameter) that ranges from 1120 (Mth1743) to 3219 (Mth1692). This size effect is also reflected in the number of filled levels (NL parameter) or occupied molecular orbitals (MOs) which is larger than 4000 for Mth1692. The proteins are characterized by different net charges, positive (+3) for Mth0637, neutral for Mth1880, and negative for the others. In general, regardless of whether the protein bears or not a net charge, there will be a number of counterions interacting with the charged groups of the protein. These counterions, however, do not significantly alter the magnitude and direction of the dipole moment vector of the protein as shown by the author in recent quantum mechanical study of Charybdotoxin, a small venom peptide that interacts with the Kcsa potassium channel [24].

Table 1. Number of atoms (NA), number of filled levels (NL), net charge, and dipole moment (in Debye) of the six Mth proteins.

| Protein | PDB id | NA | NL | Charge | $\mu$(Debye) |
|---------|--------|------|------|--------|--------------|
| Mth0637 | 1JRM   | 1692 | 2321 | +3     | 646          |
| Mth0865 | 1IIO   | 1204 | 1695 | −6     | 423          |
| Mth0895 | 1ILO   | 1207 | 1653 | −2     | 334          |
| Mth1692 | 1JCU   | 3219 | 4438 | −4     | 704          |
| Mth1743 | 1RYJ   | 1120 | 1530 | −5     | 290          |
| Mth1880 | 1IQS   | 1429 | 1962 | 0      | 434          |

With the above considerations in mind, we selected the dipole moment as an index for characterizing the electronic structure of six proteins of the Mth proteome. The computed dipole moment vectors are shown in Figure 2. The dipole moment arises from the spatial distribution of charged residues within each protein as well as the contribution of the dipoles associated to each peptide linkage (~3.5 Debye) [25]. The presence of a dipole in a protein indicates that the electronic charge is polarized (the arrow's tip represent the positive side of the dipole vector). This is important for biological recognition.

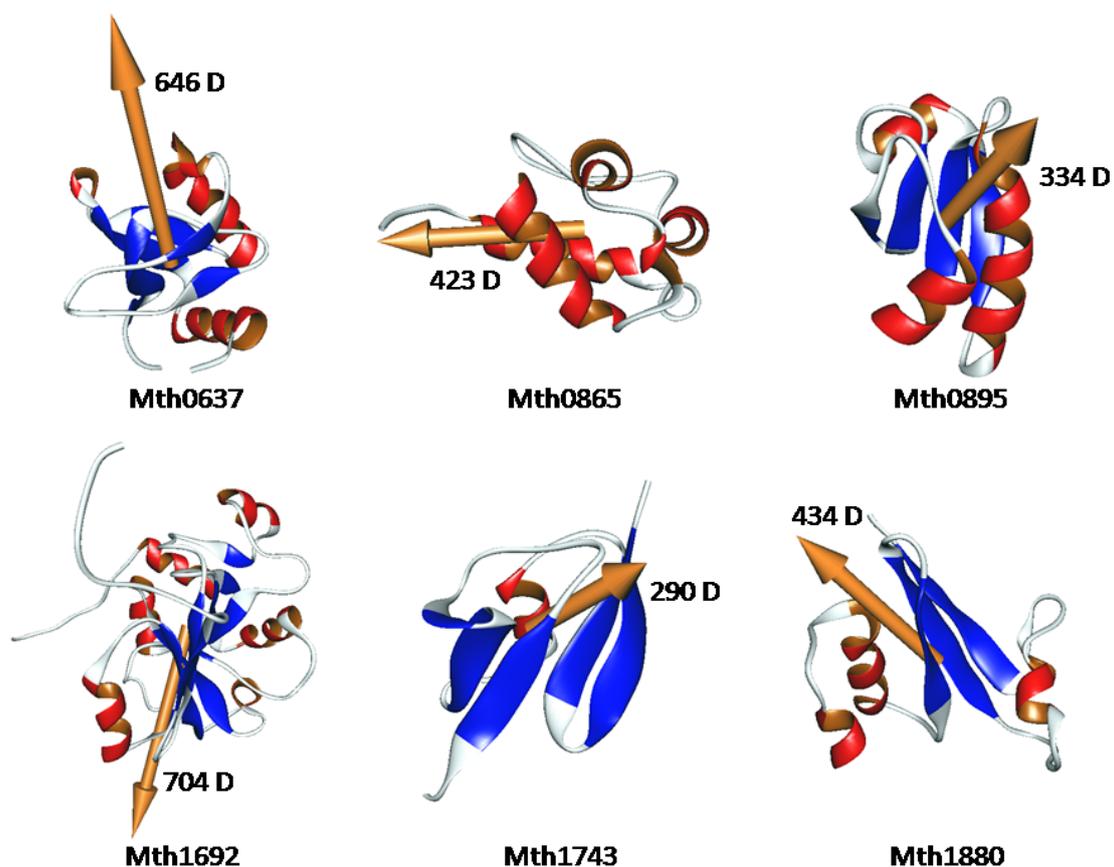

**Figure 2.** The computed dipole moment vectors of six Mth proteins. The magnitude of the dipole moment vector is given in Debye and the base of the arrow is located on the center of mass of each protein (the arrow's tip corresponds to the positive side of the dipole moment vector). The proteins are oriented as in the PDB.

From the present results it is clear that each protein possesses a characteristic electronic structure which, for convenience, can be identified using the magnitude (and direction) of its dipole moment vector. Because the electronic structure determines the biological function of a protein, a link between molecular structure and electronic structure can therefore be established, as shown in Figure 3. Dynamics is another important aspect of protein physics which contribute to establish the biological function of a protein. The conformational properties of proteins are generally investigated with the aid of classical molecular dynamics (MD) simulations [26]. Hence, the combination of electronic structure calculations and MD simulations is likely to shed new light on the biological function of proteins.

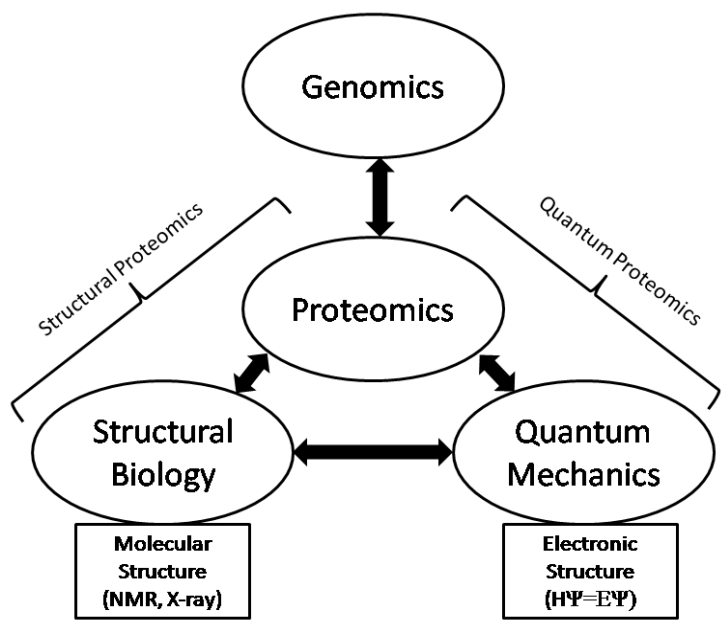

**Figure 3.** Quantum proteomics is an interdisciplinary field at the interface between proteomics and quantum mechanics.

## Conclusions

In this paper we put forward the idea of establishing a novel interdisciplinary field called *quantum proteomics* which is placed at the interface between proteomics and quantum mechanics (Figure 3). By using the experimentally-determined (NMR or X-ray) molecular structures of proteins that belong to any proteome, one can carry out quantum mechanical calculations to determine the electronic structure of such proteins. Because any protein is characterized by its unique electronic structure, it is possible to classify proteins on the basis of their electronic structural characteristics (dipole moment, energy levels, etc.) and subsequently connect these properties to biological function. Here, as an example, we selected six protein from the Mth proteome and showed that it is possible to classify or characterize them on the basis of their permanent dipole moments. We expect that in the near future a database of dipole moments and other electronic structure properties of proteins that form a proteome can be constructed and utilized in combination with other type of (structural, kinetic, biochemical, etc.) data so as to shed new light on the biological processes (e.g. interactions, enzymatic reactions, cell signaling, etc.) that emerge from such proteome.

## Acknowledgments

I thank Dr. J.J.P. Stewart for the continuous developments of the MOPAC2009 software package which allows one to investigate the electronic structure of any protein of known 3D structure. This work is supported by the Global COE program "International Center of Research and Education for Molecular Complex Chemistry" (IREMC) and the Graduate School of Engineering of Tohoku University. The *Quantum Proteomics Initiative* (QPI) is a personal scientific initiative of the author (F.P.) which represents a seed for future large-scale quantum mechanical studies of proteins that are associated to specific proteomes.## References

[1] M.J. Hubbard, Functional proteomics: The goalposts are moving, Proteomics 2002, 2, 1069–1078.

[2] A. Panchaud, M. Affolter, P. Moreillon, M. Kussmann, Experimental and computational approaches to quantitative proteomics: Status quo and outlook, J. of Proteomics 71 (2008) 19.

[3] C.H. Ahrens, E. Brunner, K. Basler, Quantitative proteomics: A central technology for systems biology, J. of Proteomics 73 (2010) 820.

[4] P. Legrain et al., The Human Proteome Project: Current State and Future Direction, Mol. Cell. Proteomics (2011), in press.

[5] Protein Databank (PDB): <http://www.pdb.org/pdb/home/home.do>.

[6] J. Nelson, *Structure and function in cell signaling*, Wiley, Chicester, 2008.

[7] J. Colinge, K.L. Bennett, Introduction to computational proteomics, PLOS Comput. Biol. 3 (2007) 1151.

[8] L. Shiff, *Quantum Mechanics*, Mc-Graw Hill, New York, 1955.

[9] *Quantum Aspects of Life*, edited by D. Abbott, P.C.W. Davies and A.K. Pati, London, Imperial College Press, 2008.10